# Magnetic braking of T Tauri stars

P.J. Armitage[1] and C.J. Clarke[2] ⋆
[1] *Institute of Astronomy, Madingley Road, Cambridge, CB3 0HA*
[2] *School of Mathematical Sciences, Queen Mary & Westfield College, Mile End Road, London, E1 4NS*

**ABSTRACT**

We construct models for the rotation rates of T Tauri stars whose spin is regulated by magnetic linkage between the star and a surrounding accretion disc. Our models utilise a time-dependent disc code to follow the accretion process and include the effects of pre-main-sequence stellar evolution. We find that the initial disc mass controls the evolution of the star-disc system. For sufficiently massive discs, a stellar field of $\sim 1$ kG is able to regulate the spin rate to the observed values during the classical T Tauri phase. The field then acts to expel the disc and the star spins up at constant angular momentum as a weak-line system. Lower mass discs are ejected at an early epoch and fail to brake the star significantly. We extend the model to close binary systems, and find that the removal of angular momentum from the disc by the secondary significantly prolongs the inner disc lifetime. Such systems should therefore be relatively slow rotators. We also discuss the implications of our model for the spectral energy distributions and variability of T Tauri stars.

## 1 INTRODUCTION

Measurements of the rotation rates of young low-mass stars show that they rotate slowly, with angular velocities typically an order of magnitude below the breakup speed (see e.g. Bouvier 1991 and references therein). This provides a puzzle for theories of star formation, as the combination of pre-main-sequence contraction and accretion of high angular momentum matter from a circumstellar disc would be expected to spinup stars to close to their breakup velocity. Moreover the classical T Tauri stars (in which there is strong evidence for active disc accretion) rotate *slower* as a class than the weak-line systems in which such evidence is lacking (Edwards et al. 1993; Bouvier et al. 1993, 1995). As the primary distinction between classical and weak-line T Tauri stars (henceforth CTTs and WTTs) is the presence or absence of disc accretion, these results suggest that the unknown angular momentum loss mechanism is more efficient in those systems possessing discs.

A model which succeeds in coupling angular momentum loss to the presence of disc accretion assumes that T Tauri stars possess ordered magnetic fields that link the star to its disc. The field lines are twisted due to differential rotation between the star and the disc, and exert a torque on both the star and the disc material. If the field is strong enough, these torques act to disrupt the inner regions of the disc at some magnetospheric radius $R_m$ that depends on the accretion rate and field strength (Ghosh & Lamb 1979; Königl 1991). If this radius is beyond the corotation radius (where the Keplerian angular velocity in the disc equals that of the star) then accretion is cut off altogether (Clarke et al. 1995); otherwise matter flows in along the field lines to strike the star with an accretion column geometry. The available spectroscopic and photometric evidence points to magnetospheric rather than boundary layer accretion being common in T Tauri stars (Hartmann 1994).

A number of authors have modelled the evolution of stellar rotation rates using such a model (Cameron & Campbell 1993; Yi 1994; Ghosh 1995; Cameron, Campbell & Quaintrell 1995). Although their assumed magnetic field configurations differ widely, these authors concur that a dipole magnetic field of a few $\times 100 - 1000$ G can regulate the stellar spin to the observed values. No CTT has a directly measured magnetic field, but such values are consistent both with the strong X-ray activity of T Tauri stars (Montmerle et al. 1992) and the measured field of the WTT TAP 35 (Basri, Marcy & Valenti 1992).

Previous studies have assumed that the accretion rate onto the star $\dot{M}$ can be prescribed as a function of time, and have then calculated the steady-state magnetospheric radius based on $\dot{M}$ and the stellar parameters. In this paper we relax that assumption, and use a time-dependent disc code to follow the evolution of the disc in response to the combined effects of internal viscous and stellar magnetic torques. This permits us to treat consistently the regime where the magnetospheric radius lies beyond corotation–in this regime accretion is cut off and no steady-state solution for $R_{\rm m}$ exists. We also investigate the influence of close binary companions on the stellar spin using our model. It is well known that in *very* close systems (with orbital periods of $\sim$ a few days) tidal effects can induce synchronous rotation between the orbital and spin periods (Tassoul & Tassoul 1992). The resulting spin periods would be faster than average for CTTs and most WTTs. Recent X-ray studies of the Hyades cluster (in which the X-ray flux is believed to be an indicator of stellar rotation) demonstrate that stars in binaries have greater X-ray luminosities than similar single stars, with the strongest effect at the small separations where tidal synchronization would be expected to operate (Pye et al. 1994; Stern, Schmitt & Kahabka 1995). Here we explore whether wider binaries, with separations of a few a.u., can exert a less direct influence on stellar rotation via tidal truncation of the accretion disc. This may modify the disc lifetime and so change the efficacy of magnetic disc braking.

---

⋆ Present address: Institute of Astronomy, Madingley Road, Cambridge, CB3 0HA.

The outline of this paper is as follows. Section 2 describes the models for the star, disc and star-disc linkage that we use in our calculations. The results of the model for the rotation of single and binary stars are presented in Section 3. The infra-red spectral energy distributions expected in our model are considered in Section 4. Section 5 discusses these results and summarizes our conclusions.

## 2 DESCRIPTION OF THE MODEL

### 2.1 Star-disc magnetic linkage

We model the unperturbed (in the absence of the disc) stellar magnetic field as a dipole whose axis is aligned with the spin axis of the star and perpendicular to the disc plane. The assumption of alignment is made for simplicity only: in real systems the detection of hotspots suggests that close to the stellar photosphere the accretion flow departs from axisymmetry. Away from the inner edge, however, the flow through the disc should be largely axisymmetric even in the non-aligned case, as a consequence of the viscous inflow timescale greatly exceeding the dynamical. Models calculated assuming alignment should therefore apply even to systems that are non-aligned. The assumption of a field with a strong dipolar component *is* crucial–we discuss the evidence for such a large-scale field in Section 5. With these assumptions, the vertical component of the magnetic field at distance $R$ from the star is given by

$$B_z = B_* \left(\frac{R}{R_*}\right)^{-3}, \tag{1}$$

where $R_*$ is the stellar radius and $B_* R_*^3 = \mu$, the stellar dipole moment. The magnitude of the magnetic torque felt by an annulus of the disc of width $\Delta R$ is then

$$T_B = B_z B_\phi R^2 \Delta R, \tag{2}$$

where $B_\phi$ is the toroidal component of the field generated by the shearing of field lines anchored both to the star and to the disc.

Evaluation of $B_\phi$ requires an assumption as to what mechanism first acts to limit the growth of toroidal field. We follow Livio & Pringle (1992) and assume that this is reconnection of twisted field lines in the magnetosphere, leading to an equilibrium toroidal field

$$B_\phi = B_z [\Omega(R) - \Omega_*]/\Omega(R), \tag{3}$$

where $\Omega$ is the angular velocity of the Keplerian disc and $\Omega_*$ that of the star. Apart from near corotation (where $B_\phi$ vanishes) the field lines are thus twisted until $B_\phi \sim B_z$. We note that our assumptions yield a twist in the magnetic field lines that is relatively conservative compared to some other authors (Campbell 1992), and thus we require stronger stellar fields for the same torque at the disc surface. In the absence of measured fields in CTTs this amounts primarily to changing the arbitrary normalisation of the stellar field strength. The radial dependence of the twist also depends on the assumptions made; however as the fall-off of the torque is dominated by the dipolar nature of the field we do not expect the exact form of equation (3) to be important.

The above analysis assumes that the field lines linking the disc outside corotation remain closed, and thus able to transmit a spin-down torque to the star. This is consistent with our assumption of a small twist in the field lines, and requires that the magnetic diffusivity in the disc material be relatively large. If, conversely, the field is twisted to a larger pitch angle, then it is prone to a catastrophic breakdown, in which the previously closed magnetic field configuration becomes open (Lynden-Bell & Boily 1994). The resultant magnetosphere has an inner part where the field lines are closed, and an outer region where the field lines threading the disc are open. Models with a magnetic field configuration of this general character have been proposed by several authors (Lovelace, Romanova & Bisnovatyi-Kogan 1995; Ostriker & Shu 1995), and have the attractive property that the open field lines may be able to power the outflows that are observed in some young stars. The results of these investigations suggest that spin regulation *is* still possible within this field geometry, although more complex intermittent behaviour also occurs. A combination of these magnetic field models with time-dependent disc codes will be required to further quantify how such models may differ from the generic closed magnetosphere picture used here.

### 2.2 Disc evolution

We follow the evolution of the disc surface density $\Sigma(R,t)$ by solving for the combined response of the disc acted on by internal viscous and external magnetic torques. With a magnetic torque given by equations (1) - (3) the resulting equation for the disc surface density evolution is (Livio & Pringle 1992),

$$\frac{\partial \Sigma}{\partial t} = \frac{3}{R}\frac{\partial}{\partial R}\left[R^{1/2}\frac{\partial}{\partial R}(\nu \Sigma R^{1/2})\right] + \frac{1}{R}\frac{\partial}{\partial R}\left(\frac{\Omega - \Omega_*}{\Omega}\frac{B_z^2 R^{5/2}}{\pi\sqrt{GM_*}}\right), \tag{4}$$

where $B_z$ is given by equation (1) and $\nu$ is the kinematic viscosity in the disc.

We adopt a form for $\nu$ based on a fit to the calculations of disc structure by Bell & Lin (1994). These assume the standard 'alpha' prescription, in which the viscosity is parameterized in terms of the local disc sound speed $c_s$ and scale height $H$ via $\nu = \alpha c_s H$, with $\alpha$ a dimensionless parameter. A power law fit to their thermal equilibrium curves gives

$$\nu = 0.3 \alpha^{1.05} \Sigma^{0.3} R^{1.25}. \qquad (5)$$

This expression provides a reasonable fit to the computed curves for input mass fluxes around $10^{-(7-8)} M_\odot \mathrm{yr}^{-1}$ and radii $R \lesssim 0.5$ a.u. It is unlikely to be a good approximation for larger radii and/or smaller mass fluxes where the temperature (and thus opacity) is very different. As the form of the viscosity is subject to considerable uncertainty in these regimes we have chosen not to attempt to construct a more complex expression for $\nu$ that takes into account these concerns. Rather we use equation (5) for all radii and surface densities in the disc. We then investigate how this choice affects our results by running models in which the viscosity law is varied either throughout the disc or solely in the outer regions.

The value of $\alpha$ appropriate to T Tauri discs is uncertain. Arguments based on the duration of FU Orionis outbursts (assuming these to be due to disc thermal instabilities) imply a low value (Clarke, Lin & Pringle 1990; Kawazoe & Mineshige 1993; Bell & Lin 1994). We therefore take $\alpha = 10^{-3}$ for our models.

### 2.3 Stellar model

The stellar model used is a $0.9985\, M_\odot$ model calculated using the most recent version of the Eggleton code (Pols et al. 1995), as modified for pre-main-sequence evolution by Tout (private communication). For spin calculations we require only the time-dependence of the stellar radius $R_*$ and moment of inertia $I_* = k^2 M_* R_*^2$, where $k$ is the dimensionless radius of gyration. The evolution of $R_*$ and $k^2$ for the model star is shown in Fig. 1. The star contracts from approximately 5 $R_\odot$ to 1 $R_\odot$ in the course of $\approx 3 \times 10^7$ yr, implying a decrease in $I_*$ by a factor of $\sim 25$ as a result of contraction. A further factor of two decrease in $I_*$ is due to the changing internal structure of the star. While the star is completely convective $k^2$ remains nearly constant at $k^2 \approx 0.2$, a value that declines between $10^7$ yr and $3 \times 10^7$ yr to $\approx 0.1$ as a radiative core develops. We argue later that this change in structure may be important in understanding the rapid spin of older WTTs.

The magnetic torques that lead to spin-down are applied to the star at the surface and at high latitude. Differential rotation may then occur at varying depths and latitudes, depending on the efficiency of angular momentum transport within the star. The strong magnetic fields within our stars will assist in enforcing rigid rotation, we therefore make the simple assumption of solid body rotation throughout the star at angular velocity $\Omega_*$. The equation for the spin evolution is then

$$\frac{d\Omega_*}{dt} = \frac{1}{I_*}\left(T - \Omega_* \frac{dI_*}{dt}\right), \qquad (6)$$

where $T$ is the net external torque on the star, comprising both a magnetic linkage contribution $T_\mathrm{mag}$ and a torque due to accretion of disc matter $T_\mathrm{acc}$. Integrating equation (2) from the inner edge of the disc at $R_m$ to infinity gives an expression for the magnetic torque on the star due to linkage to the disc,

$$T_\mathrm{mag} = \frac{\mu^2}{3}\left(R_m^{-3} - 2 R_c^{-3/2} R_m^{-3/2}\right), \qquad (7)$$

where $R_c$ is the corotation radius $R_c = (GM_*/\Omega_*^2)^{1/3}$. The magnetic linkage will therefore lead to a spin-down torque on the star if the magnetospheric radius is large enough relative to the corotation radius,

$$\left(\frac{R_m}{R_c}\right) > 2^{-2/3} \approx 0.63. \qquad (8)$$

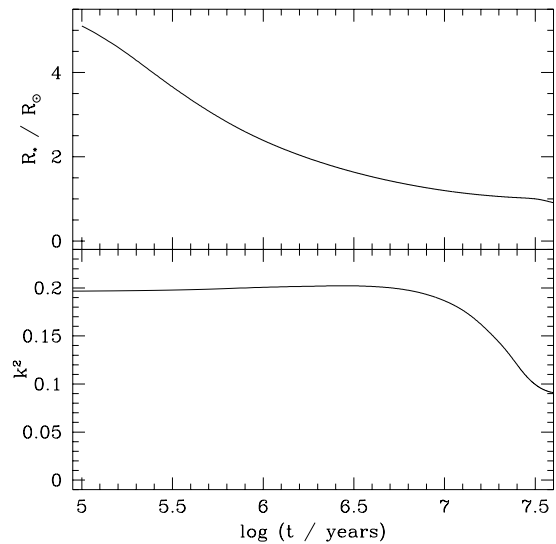

**Figure 1.** Evolution of the stellar radius $R_*$ and squared radius of gyration $k^2$ for the $0.9985\, M_\odot$ stellar model.

The accretion torque $T_\mathrm{acc}$ is given by the product of the accretion rate onto the star and the specific angular momentum of disc material at the inner edge of the disc. Strictly, the inner edge of the disc will correspond to that radius where the magnetic torques have enforced corotation of disc matter with the star, so that the spin-up torque due to accretion is $\dot{M} R_m^2 \Omega_*$. Numerically, however, we are unable to use this form, as our code assumes Keplerian disc rotation and does not resolve the narrow region of the disc where departures from Keplerian rotation occur. We therefore define $R_m$ to be the radius of the first occupied zone in our code (where Keplerian rotation still exists), and calculate

$$T_\mathrm{acc} = \dot{M} R_m^2 \Omega, \qquad (9)$$

using the Keplerian value for $\Omega$ at radius $R_m$. As the magnetic field drives strong inflow through $R_m$, essentially all the angular momentum of disc matter passing this radius will be transferred to the star. Hence, although the balance between $T_\mathrm{mag}$ and $T_\mathrm{acc}$ will be different, our method will give the correct net torque $T$ exerted on the star.

We assume that the stellar dipole field is generated via a dynamo mechanism, and take a simple linear form for the scaling of the stellar surface field with rotation rate. Specifically

$$B_* = B_0 \left(\frac{P}{4\mathrm{dy}}\right)^{-1} \quad (10)$$

with $B_0$ the normalisation for a stellar spin period $P$ of 4 dy. This is the form expected for a dynamo that scales with the Rossby number, except that we have not included effects due to changes in the convective turnover time during the pre-main-sequence (Gilliland 1986).

### 2.4 Numerical method and boundary conditions

The disc evolution equation (4) is integrated on an Eulerian radial grid evenly spaced in $\sqrt{R}$, using a simple first-order explicit scheme for the diffusive term and Lelevier differencing (Potter 1973) for the magnetic torque term. To follow the spin evolution we calculate the net stellar torque $T$ and $\mathrm{d}I_*/\mathrm{d}t$ at each timestep. For the latter we use

$$\frac{\mathrm{d}I_*}{\mathrm{d}t} = M_* \frac{\mathrm{d}}{\mathrm{d}t}\left(k^2 R_*^2\right) + \dot{M} k^2 R_*^2, \quad (11)$$

and calculate the first term on the right-hand side using cubic spline interpolation from the tabulated stellar model output. The accretion rate is therefore incorporated into the calculation of $\mathrm{d}I_*/\mathrm{d}t$ and used to update $M_*$, though in every other regard coupling between the accretion process and the stellar model is ignored. In particular, we do not allow for adiabatic shrinkage of the star as a consequence of accretion, an effect that would be important at high accretion rates. Accordingly, we commence our calculations at relatively low accretion rates, where the assumption that the stellar evolution proceeds independently from the accretion should be valid.

At the inner boundary we allow free flow through the inner edge if $R_m < R_c$. If $R_m > R_c$ the boundary condition is set to enforce no accretion onto the star, as described in a previous paper (Armitage 1995). For our simulations of discs around single stars we set $\Sigma = 0$ at the outer disc boundary $R_{\mathrm{out}}$, this corresponds to a 'no-torque' boundary condition and allows free flow of mass across the outer edge.

For binaries, we consider systems that have separations of a few a.u., these systems being comfortably wider than those in which tidal synchronization is important. On the main sequence, binaries of this separation typically have low ($e \sim 0.3$ or less) eccentricity (Duquennoy & Mayor 1991). We assume that this is also broadly correct for the pre-main-sequence binaries we are considering here. The influence of the companion on the accretion disc will then be twofold. Firstly, the disc will be truncated (at a radius typically $\sim 0.4 \times$ the pericentre distance, $R_{\mathrm{peri}}$), with further infall most probably accumulating in a circumbinary, rather than circumstellar, disc. Second, the companion acts to remove angular momentum from the outer regions of the disc, thereby preventing the disc swelling beyond the truncation radius.

To represent these effects, we set up our binary runs with a disc extending out only as far as $0.4 \times R_{\mathrm{peri}}$. The tidal influence of the companion is modelled by setting the radial velocity at the outer disc edge to zero. This 'brick-wall' boundary condition is the same as that typically employed in dwarf nova disc modelling.

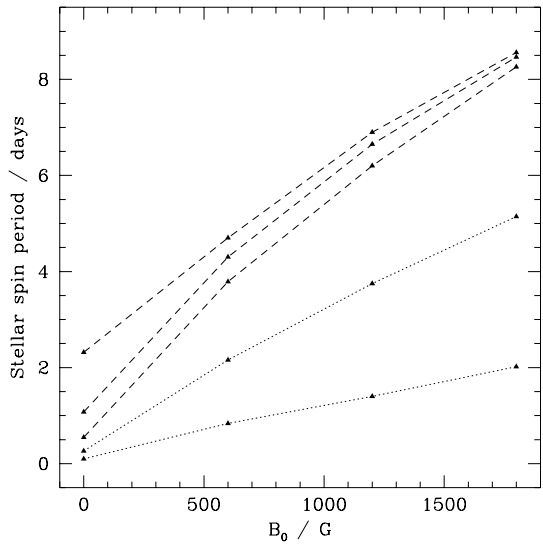

**Figure 2.** Stellar spin period at different ages, calculated as a function of the parameter $B_0$. From top to bottom, lines are plotted at $3 \times 10^5$, $10^6$, $3 \times 10^6$, $10^7$, and $3 \times 10^7$ yr. For the first three ages the model stars would appear as CTTs, the older models would appear as WTTs.

## 3 SPIN EVOLUTION

### 3.1 Initial conditions

We commence our calculations when the star has a radius $R = 5 R_\odot$, and take this time as $t = 0$. As we have argued earlier, we cannot expect to model the regime where the accretion rate is so high that it influences the stellar evolution, nor that where the star accretes a substantial fraction of its mass during the course of the calculation. We therefore set up our initial disc profile as a steady-state non-magnetic disc, with a viscous accretion rate that is constant with radius and set at $\dot{M} = 10^{-7} M_\odot \, \mathrm{yr}^{-1}$. To avoid a sudden burst of accretion as the magnetic field clears out the inner regions at the start of a run, the disc inner edge is initially set just beyond corotation. The disc surface density profile is extended outward until the total disc mass equals a specified $M_{\mathrm{disc}}$, at larger radii there is initially no mass. We assume that there is no mass infall onto the disc, and set the outer boundary condition typically at $R_{\mathrm{out}} = 4000 R_\odot \sim 20$ a.u.

The parameter that controls the strength of the magnetic braking is $B_0$, the normalisation constant in the dynamo relation. Ideally perhaps this could be fixed from the measured field and spin period of TAP 35. As the dynamo prescription employed here is undoubtably oversimplified, we instead treat $B_0$ as a free parameter and adjust it to match the observed rotation rates of CTTs. The results of such a calculation are shown as Fig. 2, in which the stellar spin is plotted at various ages as a function of $B_0$. All models have an initial rotation period of 8 dy, with a disc of mass $0.1 M_\odot$ whose outer boundary is at 10 a.u. For completeness we show a $B_0 = 0$ model, although by the latest times shown in the figure this model had spun up to break-up velocity and so was not being modelled correctly. At the first three isochrones shown ($t \leq 3 \times 10^6$ yr) the models were all accreting, and so those systems would appear as CTTs. From the

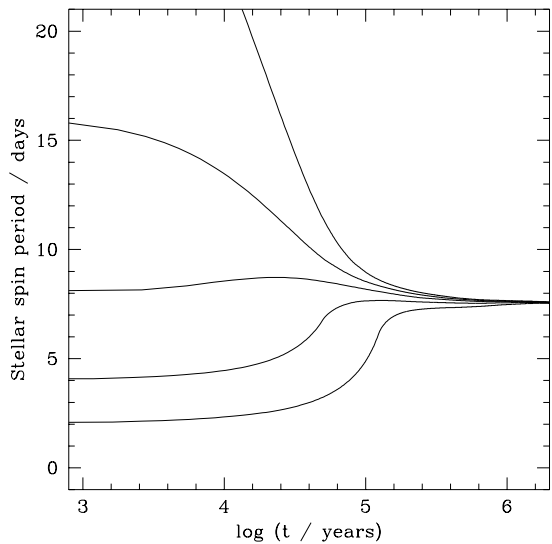

**Figure 3.** The evolution of rotation period for models with differing initial rotation period, here taken in the range 2-32 dy.

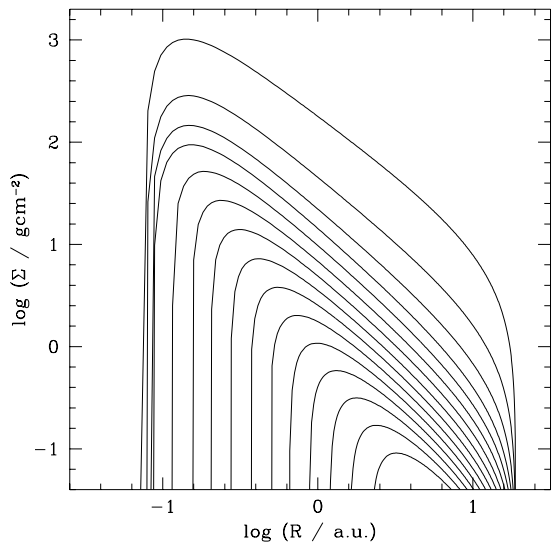

**Figure 4.** Evolution of the disc surface density profile for the model with $M_{\rm disc} = 0.1 M_\odot$, and outer boundary at 20 a.u. Curves are plotted at intervals of $2 \times 10^6$ yr, from $2 \times 10^6 - 3 \times 10^7$ yr.

figure, it can be seen that to reproduce CTTs' rotation periods of $\sim 7$ dy requires $B_0 \approx 1500$ G, and so we adopt this value in all subsequent calculations. We note immediately that with this choice of $B_0$, the spin period at later times ($10^7 - 3 \times 10^7$ yr) lies in the 2-5 dy range, similar to typical WTTs. At these epochs the models have indeed ceased accreting, and so would appear as weak-lined systems.

Previous authors have argued that magnetic braking acts to bring the star to a quasi-equilibrium spin period, in which the net torque from disc linkage and accretion is small. Using our code, we have investigated the timescale on which this quasi-equilibrium is achieved. Fig. 3 shows the effect of magnetic linkage on models with differing initial rotation periods (2-32 dy). The same disc surface density profile (with an inner edge set just beyond $R_c$ for the slowest rotating model) was used for all the runs. The slow rotators have initially weak fields and hence spin-up rapidly, while conversely the rapid rotators experience immediate magnetic braking as a result of their strong fields. We find that all models are brought to a common rotation period (here 7-8 dy) on a timescale of $\sim 10^5$ yr. This agrees well with analytic estimates of the magnetic braking timescale (Clarke et al. 1995).

This timescale is significant in two respects. Firstly, as the lifetime of CTTs is much greater than $10^5$ yr, it implies that provided magnetic braking is operating the system rapidly loses any 'memory' of its initial angular momentum. We can thus compute models assuming an arbitrary initial spin period (8 dy), and be confident that this choice does not control the later evolution. Second, it is relatively *long* compared to, for example, the expected intervals between FU Orionis outbursts in some models (Bell & Lin 1994). In the absence of other spin-down mechanisms the earliest stages of disc accretion (where $\dot{M}$ is high enough to overwhelm the stellar field) might well lead to rapid rotation. The timescale of $10^5$ yr for magnetic braking then implies that we *would* expect to observe some rapidly rotating CTTs spinning down from that early phase. During interoutburst T Tauri stars susceptible to FU Orionis events would likewise be expected to spin relatively rapidly.

### 3.2 Single stars

The evolution of the disc surface density profile with time is shown in Fig. 4, with profiles plotted at equal time intervals of $2 \times 10^6$ yr. The model had an initial disc mass of 0.1 $M_\odot$ and a zero-torque outer boundary at 20 a.u. For this model the magnetospheric radius remains approximately constant and close to the corotation radius for the first $8 \times 10^6$ yr. This is despite the fact that the accretion rate falls off rapidly, decreasing by an order of magnitude on a timescale of $\sim 2 \times 10^6$ yr. The magnetospheric radius remains at corotation even at low accretion rates because more matter can flow in from larger radii to 'reinforce' the disc and overcome the magnetic torques. This continues until the disc becomes globally weak, so that there is no longer sufficient mass at large radii. After this time, the field acts to expel the disc material, and pushes $R_{\rm m}$ rapidly outwards from corotation.

The time-dependence of the stellar rotation period corresponding to this model is shown in Fig. 5, together with four models computed with lower initial disc masses ($10^{-3} M_\odot$, $3 \times 10^{-3} M_\odot$, $10^{-2} M_\odot$, and $3 \times 10^{-2} M_\odot$). All four models with the highest disc masses display similar behaviour. While the magnetosphere is close to corotation the rotation period is regulated in the range between 6 and 10 dy. During this period the model stars are accreting, and thus would appear to be CTTs (at least as regards IR excess—as we have already noted the accretion rate itself is very low in the latter half of this phase). Amongst these four models the systems with the more massive discs rotate faster while they are CTTs, this is because higher disc mass implies a larger $\dot{M}$ and thus a larger accretion torque. These models all expel their discs shortly before $10^7$ yr, after this they spin-up by approximately a factor of two due to their vestigial stellar evolution. In common with previous authors (Cameron, Campbell & Quaintrell 1995), we find that the final stellar rotation period is independent of the initial disc mass provided that it is high enough for magnetic braking to operate. The systems with the highest disc masses initially spinup more due to accretion, but this is compensated by longer disc lifetime (and hence a prolonged time over which braking can occur).

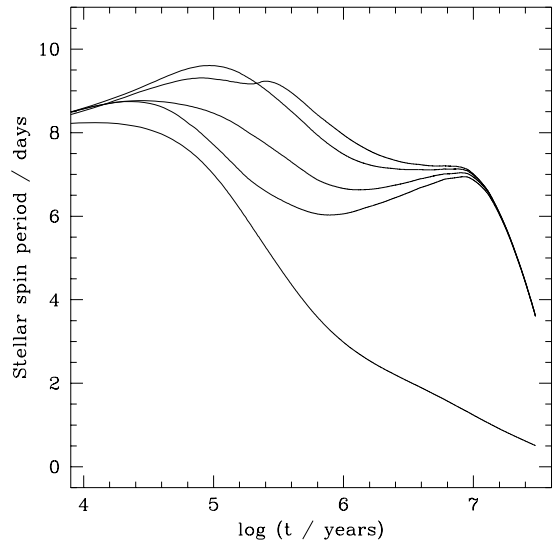

**Figure 5.** Time-dependence of the stellar spin period for models computed with different initial disc masses. In all cases the outer boundary was appropriate for a 'single' star and was set at 20 a.u. From top downwards at $t = 10^6$ yr the curves represent $M_{\rm disc} = 3 \times 10^{-3}, 10^{-2}, 3 \times 10^{-2}, 10^{-1}$ and $10^{-3} M_\odot$ respectively.

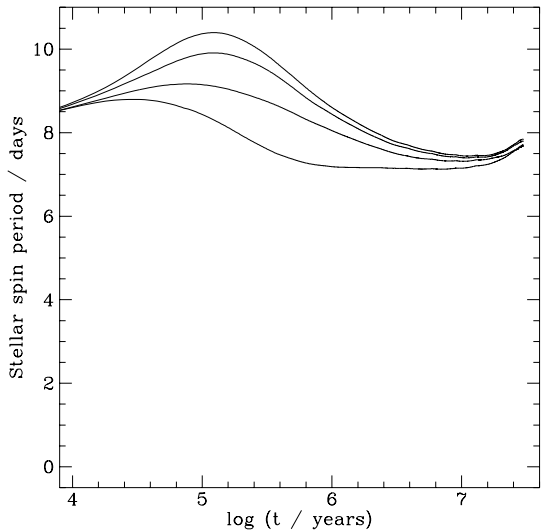

**Figure 6.** Time-dependence of the stellar spin period for models computed with an outer boundary condition appropriate to a binary system. From top downwards the models assume $R_{\rm peri} = 1$, 2, 4, and 8 a.u.

For the lowest mass disc the behaviour is different. In this case the magnetic torques expel the disc material to large radii almost immediately, from where it is unable to brake the star effectively. The star then spins up as it evolves along its pre-main-sequence track. We note that for such systems the assumed initial spin period *is* important, and indeed determines the spin at all later stages in the absence of any alternative braking mechanism. However, for any reasonable initial spin the large decrease in $I_*$ (approximately a factor of 50 in our stellar model) will ensure that systems with such low disc masses will become rapid rotators. As very little matter is accreted, they would appear as WTTs from an early epoch. The critical disc mass for this to occur is obviously a function of $\alpha$ and the assumed viscosity in the disc–for the parameters assumed here it is $\sim 10^{-3} M_\odot$.

The viscosity prescription adopted for these calculations will fail at large radii in the disc, where the temperatures and opacities are very different. To explore how sensitive these results are to the form of the viscosity, we have computed models in which the power-law form of equation (5) is modified beyond a given radius (typically 1 a.u.). We find that the general behaviour shown in Fig. 5 (i.e. the shape of the curves and the dichotomy between high and low disc masses) persists in these alternative models. However, the timescale for the disc to be expelled, and thus the final spin period, does vary substantially. Assuming a higher viscosity in the outer disc leads to higher accretion rates and shorter disc lifetime, both of which imply faster rotation. Conversely, lower viscosity at large radii produces more braking. We would thus caution that with our current knowledge of disc viscosity it is not possible to predict in detail the CTTs and WTTs spin distribution. However magnetic braking via star-disc linkage is a robust mechanism in that it will produce distributions similar to those observed provided only that the disc lifetime in T Tauri stars is 5-10 Myr. In particular, it would operate successfully even if some other effect, perhaps stellar winds, dissipated the disc at the end of the CTTs phase.

### 3.3 Binary stars

For our runs intended to mimic binary systems, we again set up the disc to produce an initial accretion rate of $10^{-7} M_\odot$ yr$^{-1}$, with an inner hole just larger than $R_c$. In this case, though, we extend the disc profile out as far as $0.4 \times R_{\rm peri}$, the binary pericentre distance. This is the radius at which the disc is tidally truncated by the binary companion. The boundary condition at the outer edge is set to enforce zero radial velocity.

Fig. 6 shows the results for binaries with pericentre separations in the range 1-8 a.u. As for single stars, magnetic braking is effective in maintaining the stellar rotation rate in the 7-10 dy range for the few Myr duration of the CTT phase. The differences between the models are due to the larger disc masses in the wider binaries. The wider systems are able to maintain a relatively high accretion rate for longer than the close systems, and as a result they have shorter equilibrium spin periods.

Although there are detailed differences due to the boundary conditions, the general behaviour of the binary models is very similar to that of the single stars up to a time of $\sim 10^7$ yr. We would therefore expect similar rotation rates in CTTs, whether they were single or members of a binary system. After $10^7$ yr, however, there is an obvi-ous distinction between the binary and single star models. Whereas the single star models expel their inner discs and spinup, the binary models *retain* mass just outside corotation. This continues to provide magnetic braking, and as the accretion torque falls off with time the models actually begin to spindown. They never achieve WTTs rotation periods, although the accretion rate is negligible in these models after the first few Myr.

The continuing spindown of the binary models is evidently due to these stars being unable to expel their inner discs, and this in turn is a consequence of the tidal effects of the companion. The companion is able to extract an arbitrary amount of angular momentum from the disc material, which can therefore remain at small radii despite the continual injection of angular momentum from the action of magnetic torques. The disc is effectively an intermediary in a continuing transfer of angular momentum from the spin of the primary, via the disc, to the orbit of the secondary.

An intriguing possibility raised by this model is the possible existence in close systems of a ring of gas and dust trapped between the corotation and binary truncation radii. The material in the ring would be unable to flow inwards due to the magnetic torques near corotation, while the tidal torques from the binary companion would prevent outflow. Such a ring could exist stably for a long period, and its eventual dissipation would be controlled by processes that we have not included in our models–for example the clearing effect of stellar winds. The final stellar rotation period would then depend on the time at which the disc was dissipated and magnetic braking ceased.

As will be obvious by now, determination of the predicted final rotation rates of magnetically braked binaries is by no means straightforward. Commencing with the shortest period systems, binaries with periods of a few days are expected to be synchronized by direct star-star tidal interaction. These will appear as rapid rotators compared to the mean CTT or WTT spin periods. Somewhat wider systems, where the *entire* disc is able to be cleared by the magnetic torques, should also lose their braking mechanism and spinup to be rapid rotators. The condition for this is that the binary disc truncation radius lies inside corotation, so that the magnetic torques drive rapid inflow across the whole surface of the disc. For a typical CTT rotation period of 8 dy, this effect will persist up to pericentre distances of $\approx 0.2$ a.u. Beyond that, for binaries of separation a few a.u., we predict slow rotation due to coupling of the star to a trapped ring of material as described above. Finally, for sufficiently wide binaries, we expect rotation rates comparable to single stars, as a result of the outer disc's characteristic timescale being so long that the inner disc is effectively unaware of the outer boundary at all. For a very crude estimate of the radius where this final transition occurs, we note that the disc thermal timescale ($\alpha^{-1}\Omega^{-1}$) reaches 1 Myr at around 100 a.u. The viscous timescale, which can safely be taken to exceed the thermal by a substantial factor, is thus likely to be greater than any other timescale in the system by this radius in the disc. We therefore anticipate that wide binaries ($\sim 10^2$ a.u. or greater) should have rotation rates similar to single stars.

## 4 SPECTRAL ENERGY DISTRIBUTIONS

A prediction of the magnetic models presented in Section 3 is that Classical T Tauri stars should have magnetospheric radii that lie close to corotation, at least for the mass fluxes $\dot{M} \lesssim 10^{-7} M_\odot$ yr$^{-1}$ considered here. The presence of such 'holes' (typically with radii of 10 $R_\odot$ or more) will modify substantially the spectral energy distributions (SED) expected in the infra-red. To calculate this, we assume that each annulus of the disc is in thermal equilibrium and radiates as a blackbody with effective temperature $T_e$. Equating the work done locally by viscous and magnetic torques with radiative cooling then gives

$$2\sigma T_e^4 = \frac{9}{4}\nu\Sigma\Omega^2 + f\frac{|B_z B_\phi|}{2\pi}R\Omega, \qquad (12)$$

where $\sigma$ is the Stefan-Boltzmann constant. The first term on the right-hand side is the usual expression for the local dissipation due to viscous torques, while the second term represents the work done by the magnetic field. The factor $f$ is included to allow for our uncertainty as to what fraction of that energy goes into heating the disc material, as opposed to being deposited in the magnetosphere or in the star. We compute models based on the two extreme cases, $f = 0$ (the 'no-reheating' limit) and $f = 1$ (the 'full reheating' limit).

The magnetic term in equation (12) can be straightforwardly evaluated from equations (1) and (3), and is independent of the viscosity or surface density profile in the disc. The rate of viscous energy dissipation in the disc can also be found analytically, from the steady state solution of equation (4). Setting the inner boundary condition to be $\Sigma = 0$ at $R = R_c$ we find,

$$\nu\Sigma = \frac{\dot{M}}{3\pi}\left(1 - \sqrt{\frac{R_c}{R}}\right)$$
$$+ \frac{\mu^2}{9\pi\sqrt{GM_*}}\left(R_c^{-3}R^{-1/2} - 2R_c^{-3/2}R^{-2} + R^{-7/2}\right). \qquad (13)$$

The first term on the right-hand side of this equation is just the normal non-magnetic expression for a disc with a zero-torque inner boundary condition at $R = R_c$. The second term represents the effect of the magnetic torque, which acts to increase $\nu\Sigma$ upstream of corotation. The field is here injecting angular momentum into the disc material, creating a magnetic barrier to the accretion flow. To maintain a given $\dot{M}$ the viscosity and surface density must rise to overcome this barrier. Using the specific form for $\nu$ given in equation (5), we have verified that this formula is in good agreement with the surface density profiles computed earlier for the inner regions of the disc.

Using equations (12) and (13) we calculate the temperature profile across the disc, and from this generate the expected spectral energy distribution in the reheating and no-reheating limits. As for the steady-state non-magnetic case this is independent of the form assumed for the viscosity. The parameters required to generate a magnetic disc spectrum are then the accretion rate $\dot{M}$, the corotation radius $R_c$, and the quantity $\beta = \mu^2/(9\pi\sqrt{GM_*})$, which characterizes the strength of the magnetic effects. We compare our spectra with two non-magnetic models. Firstly (type 1 model) an undisrupted disc model in which the disc extends down to the stellar surface, which we take at $R = 3R_\odot$. Second (type 2 model) a model in which the influence of the field is included only insofar as it creates a large hole in an otherwise non-magnetic disc (i.e. we take the inner disc radius to be at corotation but set $\beta$ to zero in the above expression). Fits to observed spectra are often made using such a model (Hillenbrand, private communication), we therefore investigate how well it captures the effect of the star-disc coupling on the infra-red fluxes.

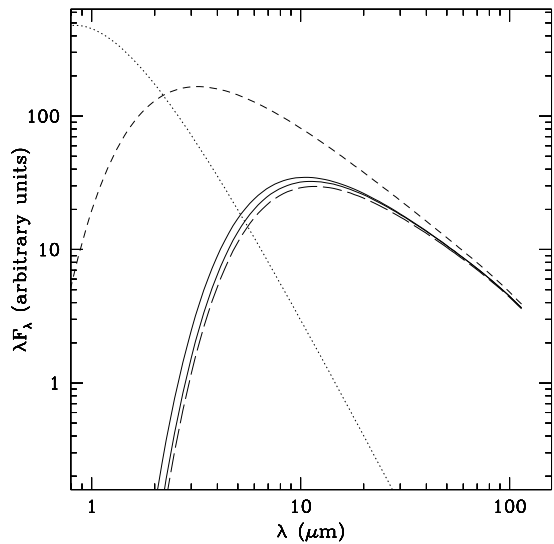

**Figure 7.** Spectral energy distributions calculated for an accretion rate of $\dot{M} = 10^{-7} M_\odot$ yr$^{-1}$, using the four models described in the text. The solid lines represent the magnetic model with either full reheating (upper curve) or no reheating (lower curve). The short dashed line shows an undisrupted disc spectrum, and the long dashed line the spectrum of a non-magnetic disc disrupted at corotation. The dotted line shows a blackbody spectrum representative of the stellar contribution to the SED.

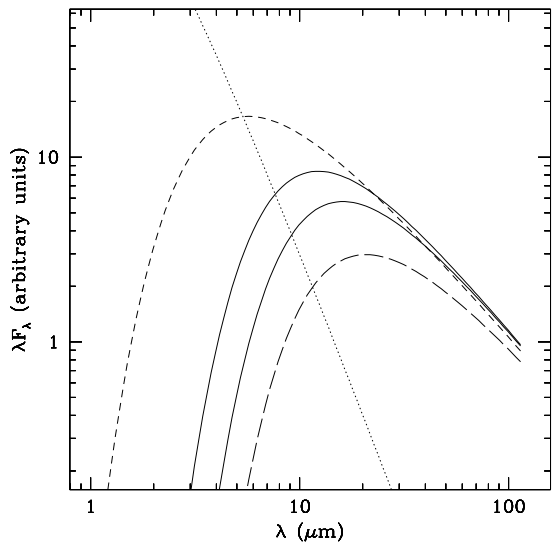

**Figure 8.** Spectral energy distributions calculated for an accretion rate of $\dot{M} = 10^{-8} M_\odot$ yr$^{-1}$, using the four models described in the text. The solid lines represent the magnetic model with either full reheating (upper curve) or no reheating (lower curve). The short dashed line shows an undisrupted disc spectrum, and the long dashed line the spectrum of a non-magnetic disc disrupted at corotation. The dotted line shows a blackbody spectrum representative of the stellar contribution to the SED.

Fig. 7 shows spectra for an accretion rate of $10^{-7} M_\odot$ yr$^{-1}$, stellar magnetic field $B_* = 750$ G, and corotation radius $R_c = 16.8 R_\odot$ (corresponding to a rotation period of 8 dy). The field strength is chosen to be consistent with the inferred dynamo relation for a system with this rotation rate. The stellar radius is taken as $R_* = 3 R_\odot$, appropriate for a relatively young $t \lesssim 10^6$ yr CTT system. For comparison we also plot a blackbody spectrum corresponding to the effective temperature and luminosity of our model star. We take a luminosity of $L_\odot$ (which is within a factor of two of the model luminosity between $10^6$ and $10^7$ yr) and an effective temperature of 4400 K. We do not include any contribution to the spectrum from the presence of accretion hotspots–over the wavelength range considered here any such contribution is likely to be small.

The figue shows that the removal of the inner disc leads to a large reduction in the $2-5\mu m$ disc fluxes, and infra-red colours that are much redder than for the corresponding disc that terminates in a boundary layer. Beyond $2\mu m$ the differences are clear even when the contaminating effect of the stellar emission is considered. For this accretion rate the no-reheating magnetic disc spectrum is close to that of a non-magnetic disc with the same inner radius. The full reheating spectrum shows more emission at short wavelengths, due to the increased temperatures reached in a narrow region of the inner disc if the work done by the field goes into heating the disc matter. However in either case the gross effect of the magnetic field is captured well by simply moving the zero-torque boundary condition of a non-magnetic disc out to the corotation radius.

For lower accretion rates, however, the influence of the field on the disc structure and spectrum is not as straightforward. Fig. 8 shows models computed with the same parameters as Fig. 7 except with $\dot{M} = 10^{-8} M_\odot$ yr$^{-1}$. As before, the no-reheating magnetic model displays greatly reduced fluxes at short wavelengths, as a result of removing the innermost regions. However, there is significantly *more* flux at all wavelengths than would be predicted from a non-magnetic model plus hole. This is because the magnetic torques are damming-up material outside of the corotation radius, leading to enhanced surface density and temperature in that region. Assuming full reheating increases the fluxes still further. Indeed for the reheating model the flux beyond $20\mu m$ marginally exceeds even that expected from an undisrupted disc. Fits to magnetic disc spectral energy distributions based on models of type 2 would therefore tend to overestimate the accretion rate and underestimate the size of the inner hole. For the parameters used here the best-fitting type 2 model has an accretion rate approximately twice that of the magnetic model, and an inner disc radius that is 10-20% too small.

If the stellar magnetic field is dynamo generated, then analogy with the Sun suggests that it might also be variable on observable timescales. Using the spectral models described above, we can predict the effect variable fields will have on the disc emission, and thereby quantify the estimates given in a previous paper (Armitage 1995). To do this, we first note that the viscous timescale in the disc is relatively long, and thus rapid changes in $B_*$ will occur while the surface density profile $\Sigma(R)$ is almost fixed. The viscous term in equation (12) is thus insensitive to rapid field changes, and variability on timescales of months to years will be driven solely by the magnetic term. The difference between the reheating and no-reheating spectra then provides an upper limit to the expected photometric variability (an upper limit because $f$ is unknown), and an indication of how that variability scales with wavelength.

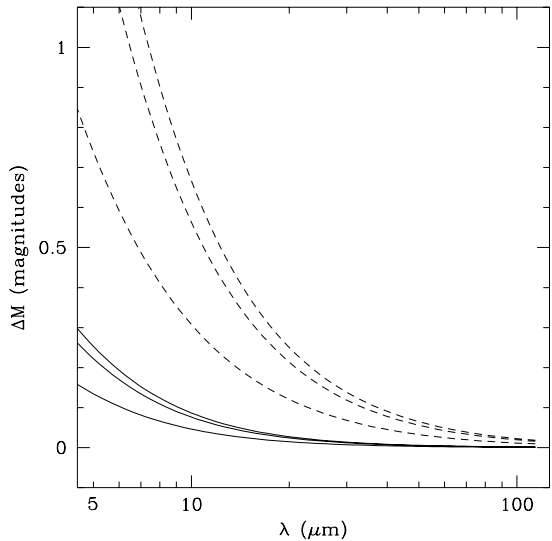

**Figure 9.** Predicted variability as a function of wavelength from rapid changes in the stellar magnetic field. The upper solid line shows the expected $\Delta M$ if the field drops from 750 G to zero, assuming full reheating and an accretion rate $\dot{M} = 10^{-7} M_\odot$ yr$^{-1}$. The lower solid curves assume the field reduces to 250 G and 500 G respectively. The dashed curves are the same, but for an accretion rate of $10^{-8} M_\odot$ yr$^{-1}$.

Fig. 9 shows the upper limits to the photometric variability $\Delta M$ at different wavelengths, computed for the $\dot{M} = 10^{-7}$ and $10^{-8} M_\odot$ yr$^{-1}$ models used earlier. For each model we show the effect of $B_*$ dropping from its initial 750 G to 500 G, 250 G, and zero. Note that these models apply for the disc flux only—at short wavelengths the presence of a dominant stellar component to the SED curtails the overall system variability. From figures 7 and 8 we expect the stellar emission to be important for $\lambda \lesssim 5\mu$m at an accretion rate of $10^{-7} M_\odot$ yr$^{-1}$, and at $\lambda \lesssim 10\mu$m at an accretion rate of $10^{-8} M_\odot$ yr$^{-1}$. The reduction in the disc IR emission at short wavelengths means that the stellar emission is important at longer wavelengths than for non-magnetic star-disc models.

From the figure, it is apparent that variability due to the action of variable magnetic torques is strongly chromatic, and falls off rapidly with increasing wavelength. For an accretion rate of $\dot{M} = 10^{-7} M_\odot$ yr$^{-1}$ $\Delta M$ could be as large as 0.3 magnitudes at 5 $\mu$m, but this upper limit falls to $< 0.1$ mag at 10 $\mu$m. Perhaps more significant than the precise numbers, we find that variability from this mechanism is expected to be greater in low $\dot{M}$ systems, because the 'background' dissipation due to viscous processes is lower in those systems. This is the case even when we consider the fact that the stellar emission is a more important contaminant for low accretion rates–the predicted variability is larger at $10\mu$m for $\dot{M} = 10^{-8} M_\odot$ yr$^{-1}$ than it is at $5\mu$m for $\dot{M} = 10^{-8} M_\odot$ yr$^{-1}$. Moreover, significant variability extends to longer wavelengths if the accretion rate is low, because the field exerts a significant influence on $T_e$ out to larger (cooler) radii. Finally, for stellar fields of similar strength and variability, $\Delta M$ is predicted to be greater in rapidly rotating systems. This is because the total viscous dissipation in the disc scales as the depth of the potential well ($\propto R^{-1}$), while the magnetic torques have a much steeper radial dependence.

## 5 DISCUSSION

We have constructed models for the braking of T Tauri stars in which the accretion disc is disrupted by an ordered stellar magnetic field. Our models differ from previous authors in that we do not assume a steady-state configuration for the disc, but instead allow it to evolve in response to viscous and magnetic torques. Compared to earlier work we find that the magnetospheric radius remains close to corotation for longer, because as the accretion rate drops matter flows in from large radius to reinforce the inner disc and maintain the inner edge at corotation. We would therefore expect braking to continue in some systems where the diagnostics of accretion (such as UV excess) were too weak to be observed. Such systems would still possess weak IR excesses and emission at mm wavelengths, though as remarked by Bouvier (1991) the discs in such systems would be optically thin at relatively small radii. We also note that a magnetosphere that lies near corotation could readily be displaced beyond $R_c$ by relatively small increases in the stellar field, thereby leading to variability in the accretion flow and luminosity (Clarke et al. 1995; Armitage 1995).

In our models, we require the parameter $B_0$ to be around 1500 G in order to reproduce the observed spin periods of T Tauri stars. This implies that the relatively slowly ($\sim 7$ dy) rotating CTTs should possess surface magnetic field strengths of 500-1000 G. This estimate is probably uncertain by a factor of two as a result of our approximate treatment of the twisted magnetic field, however it does not appear unreasonably high given the measured kG strength field in TAP 35 (though this star is a WTTs with rapid rotation period). More troubling is the fact that efficient star-disc braking of CTTs requires a field that generates significant torque at radii $R \gtrsim 10 R_\odot$. In practice this necessitates a magnetic field that has a strong dipolar component. Some support for the existence of such fields is provided by VLBI observations of HD 283447 (Feigelson et al. 1994), which indicate the presence of large-scale, ordered magnetic fields in that source. These measurements, together with the indirect evidence for magnetospheric accretion in CTTs, may therefore constrain dynamo models for field generation, some of which do favour a dipolar field geometry.

A prediction of our models is that the magnetospheric radius must lie close to corotation throughout the lifetime of CTTs. This follows simply from the rapid radial decay of a dipolar field–for the coupling outside $R_c$ to overwhelm that interior to corotation the inner disc edge must lie near the corotation radius. The observational signature of the disrupted disc, as noted by many previous authors, lies in severely depleted near-IR disc fluxes compared to models in which the disc extends to the stellar equator. This will be true regardless of the assumptions made for contentious theoretical questions such as the magnetic field configuration. However, we would caution that the details of the spectrum *are* highly model dependent, especially at low accretion rates, and important quantities such as $R_\mathrm{m}$ cannot be recovered without assuming a specific model. Detailed modelling of infra-red data, along the lines of a recent paper (Kenyon, Yi & Hartmann 1995), will be essential in discriminating between models and determining system parameters.

The mechanism described here appears to operate successfully in braking stars with a wide range of disc masses, provided that the mass exceeds some critical value. Lower mass discs are rapidly expelled by the magnetic torques, producing a runaway effect and rapid stellar spin. A population of rapid rotators *is* observed in young clusters such as α Persei (Keppens, MacGregor & Charbonneau 1995), these stars could therefore have been produced from systems in which disc braking was ineffective. A number of mechanisms could lead to such weak initial discs, including the collapse of low angular momentum clouds or encounters that disrupt the disc down to small radii.

In terms of the distinction between classical and weak-lined T Tauri stars, this work supports a twin-track model of their evolution. In this picture, some WTTs are old CTTs that have expelled their discs and are spinning up as a result of their stellar evolution–in particular the reduction in the moment of inertia when a radiative core develops. Such stars would have undergone a prolonged (5-10 Myr) disc phase. The remaining WTTs would not be simply post-CTTs, but would instead be systems which had weak discs either initially or as consequence of disruptive collisions at an early epoch. These WTTs could be young, as is suggested by the intermingling of the classes on the Hertzprung-Russell diagram. As they spun up, they would form the rapidly rotating tail of the WTTs spin distribution.

The observational correlation of stellar rotation rates with binarity offers the prospect of further testing the magnetic disc braking model. As we have noted, two distinct effects are predicted. Short period binaries (separations up to a few tenths of an a.u.) should clear away their discs entirely and spinup to rapid rotation rates. This may already have been seen indirectly in the X-ray observations of Stern, Schmitt & Kahabka (1995). Conversely, wider binaries (separations of typically a few a.u.) are predicted to be slow rotators at late times. The ongoing braking in these systems is caused by magnetic coupling to a ring of trapped material between corotation and the disc outer edge. In addition to controlling the spin, the presence of such a ring may be detectable as an IR excess in relatively old, non-accreting WTT binaries.

## ACKNOWLEDGMENTS

We are indebted to Chris Tout for making available the stellar model used in this work, and to Starlink staff for computer support. We also wish to thank Jim Pringle for many useful discussions, and the referee for suggesting a number of improvements to the paper.